\documentclass[conference]{IEEEtran}
\IEEEoverridecommandlockouts
\usepackage{cite}
\usepackage{amsmath,amssymb,amsfonts}
\usepackage{algorithmic}
\usepackage{graphicx}
\usepackage{textcomp}
\usepackage{xcolor}
\usepackage{microtype}
\usepackage{lipsum}
\usepackage{multirow}

\def\BibTeX{{\rm B\kern-.05em{\sc i\kern-.025em b}\kern-.08em
    T\kern-.1667em\lower.7ex\hbox{E}\kern-.125emX}}

\begin{document}

\title{Fault location in High Voltage Multi-terminal dc Networks Using Ensemble Learning}

\author{\IEEEauthorblockN{Timothy Flavin\IEEEauthorrefmark{2},
Bhaskar Mitra\IEEEauthorrefmark{4}\IEEEauthorrefmark{1}~\IEEEmembership{Member,~IEEE,}
Vidhyashree Nagaraju\IEEEauthorrefmark{2}~\IEEEmembership{Member,~IEEE,} and \\ Rounak Meyur\IEEEauthorrefmark{3},
\IEEEmembership{Student Member,~IEEE}}
\IEEEauthorblockA{\IEEEauthorrefmark{2}Tandy School of Computer Science, The University of Tulsa, Tulsa, OK 74104 USA}
\IEEEauthorblockA{\IEEEauthorrefmark{4}Idaho National Laboratory, Idaho Falls, ID 83402 USA}
\IEEEauthorblockA{\IEEEauthorrefmark{3}Network Systems Science and Advanced Computing, University of Virginia, Charlottesville, VA 22904 USA}
\IEEEauthorblockA{\IEEEauthorrefmark{1}Corresponding author: Bhaskar Mitra (email: bhaskarmitra1991@gmail.com)}}
%\IEEEauthorblockA{\IEEEauthorrefmark{1}Corresponding author: Bhaskar Mitra (email: bhaskar.mitra@inl.gov)}}

\twocolumn[
\begin{@twocolumnfalse}
\maketitle
\end{@twocolumnfalse}
\begin{abstract}
Precise location of faults for large distance power transmission networks is essential for faster repair and restoration process. High Voltage direct current (HVdc) networks using modular multi-level converter (MMC) technology has found its prominence for interconnected multi-terminal networks. This allows for large distance bulk power transmission at lower costs. However, they cope with the challenge of dc faults. Fast and efficient methods to isolate the network under dc faults have been widely studied and investigated.  After successful isolation, it is essential to precisely locate the fault. The post-fault voltage and current signatures are a function of multiple factors and thus accurately locating faults on a multi-terminal network is challenging. In this paper, we discuss a novel data-driven ensemble learning based approach for accurate fault location. Here we utilize the eXtreme Gradient Boosting (XGB) method for accurate fault location. The sensitivity of the proposed algorithm to measurement noise, fault location, resistance and current limiting inductance are performed on a radial three-terminal MTdc network designed in Power System Computer Aided Design (PSCAD)/Electromagnetic Transients including dc (EMTdc).
\end{abstract}

\bigskip
\bigskip]

\section{Introduction}
High Voltage Direct Current (HVdc) transmission has emerged as a leading contender for energy transmission compared to High Voltage Alternating Current (HVac) for bulk power off-shore and on-shore transmission. They are considered as a viable form of energy transportation for renewable energy transportation and interconnection of multiple grids of different frequencies. The Voltage Source Converter (VSC) design allows for the implementation of interconnected multi-terminal dc (MTdc) networks with bi-directional power flow. The Modular Multilevel Converter (MMC) has emerged as a popular choice for VSC-HVdc networks due to salient features of better scalability and higher operational efficiency \cite{mitra2018hvdc}. The remote location of HVdc networks makes it challenging for the utility crew to perform repairs in case of a disruption. Accurate location of faults on the network is thus essential for repair and restoration process. Travelling wave based methods are employed on a wide scale for accurate fault location. Mainly two methods are implemented, they are (1) single ended measurements and (2) double ended measurements \cite{Ando1985}.

Double ended measurements have traditionally provided higher accuracy of fault locations using the current and voltage measurements. This method involves setting up and maintenance of robust communication networks and expensive time synchronized devices \cite{Ying-HongLin2004}. On the contrary single ended measurements are more convenient. They are cheap but tend to provide inaccurate results as devices do not have the capability to detect the reflected peak \cite{Xu2011}. Its performance is driven by fault resistance and other network parameters. The reflected surge waves are weaker thus making their detection difficult. Inability to capture the reflected surge wave peaks would provide provide inaccurate fault locations.

Some other methods of fault location has been proposed using digital signal processing, they require device with high sampling frequency to achieve the desired results. Methods involving time-frequency analysis of the fault transients have been performed using wavelet transform \cite{Magnago1998}, implantation of such algorithms would require double ended synchronized measurements. Passive methods of fault locations have been suggested in \cite{Mohanty2016} and \cite{Mitra2021}, although their demonstration for a real-field example is yet to be demonstrated.

Through the development and advancements in the field of artificial neural networks, modern algorithms can be utilized for efficiently determining the fault location. Previously fault location through machine learning approaches has been discussed in \cite{Hanif2014}. Wavelet transforms were utilized to capture the transient data from a single terminal and the processed data was trained through a Support Vector Machine (SVM) network for fault location \cite{Hanif2014}.

In this paper, we propose a single ended fault location technique using ensemble learning method, eXtreme gradient boosting (XGB) algorithm. The single-ended post-fault voltage and current measurements are used as input to the XGB model. Several machine learning algorithms were utilized to support the selection of XGB algorithm. Results suggests that XGB achieves a consistently better accuracy compared to other algorithms despite the small folds and small sample size of the training data provided. The prediction accuracy of XGB is high and mean average error between estimating the fault location decreased from $300$ to a value closer to zero. XGB demonstrated a linear computing time with the increase in training samples.

The remainder of the paper is organised as follows: Section \ref{sec:TestSystem} includes the modelling of the MTdc system. Section \ref{sec:Method} describes the approach for fault location and reviews the ensemble learning model along with data and experimental setup. Section \ref{sec:Experiment} discusses the results and Section \ref{sec:Conclusion} concludes the paper with major findings.

\section{Test System} \label{sec:TestSystem}

\subsection{The Modular Multilevel Converter}

The modeling of the three terminal MMC in this paper is based on the design suggested in \cite{Debnath2018}. The MMC model consists of 400 half bridge sub-modules (HBSM) per arm. Hybrid discretization and relaxation algorithms described in \cite{Debnath2018} are used to define the numerical stiffness in the differential algebraic equations. The control  of  the  MMCs  is based on the strategies explained in \cite{Debnath2016}. More details about the system parameters are provided in Table \ref{Table 1}.

\begin{table}[]
\centering
\caption{System Parameters}\label{Table 1}

\begin{tabular}{|c|c|c|}
\hline
 & \textbf{Parameters} & \textbf{Value} \\ \hline
\multirow{5}{*}{\textbf{ac side}} & Voltage (L-L RMS) & 333 kV \\ \cline{2-3} 
 & Length of transmission line 1 \& 3 & 100 km \\ \cline{2-3} 
 & Length of transmission line 2 & 150 km \\ \cline{2-3} 
 & System Frequency & 60 Hz \\ \cline{2-3} 
 & Transmission line resistance & 0.03206 $\Omega$/km \\ \hline
\multirow{4}{*}{\textbf{dc side}} & Voltage (L-L) & 640 kV \\ \cline{2-3} 
 & Length of transmission line & 1000 km \\ \cline{2-3} 
 & Transmission line resistance & 0.03206 $\Omega$/km \\ \cline{2-3} 
 & MMC capacity & 1 GW  \\ \hline
\end{tabular}
\end{table}

\subsection{The MTdc Network}
\begin{figure}
    \centering
    \includegraphics[width=0.5\textwidth]{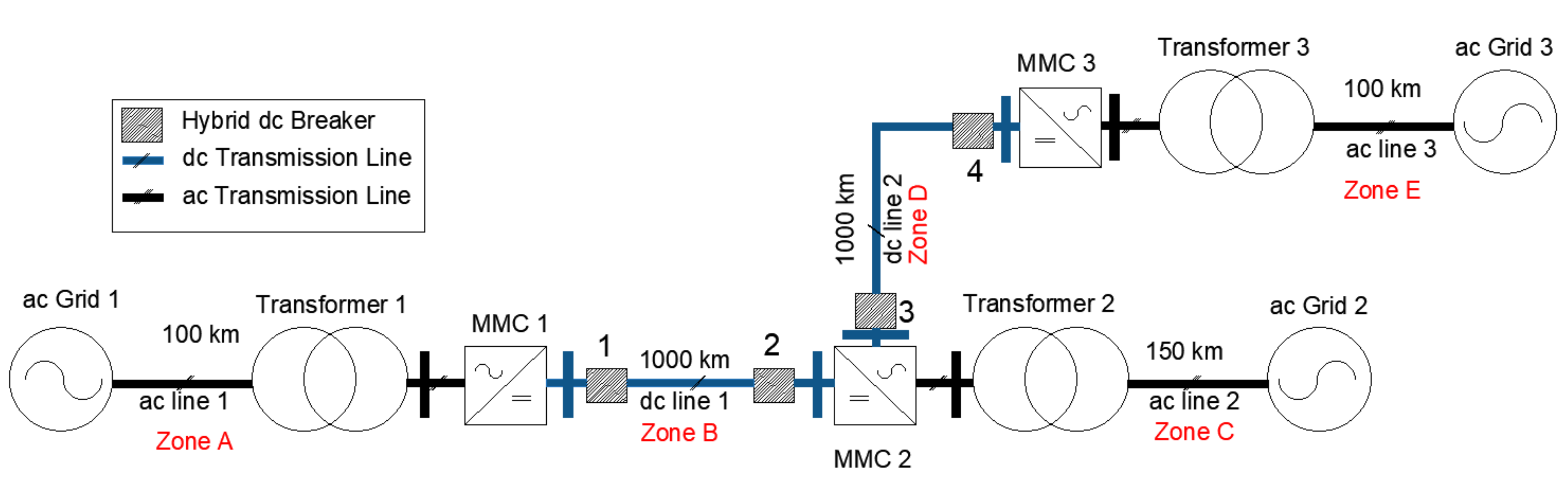}
    \caption{Radial multi-terminal dc network}
    \label{fig:zones}
\end{figure}

\begin{figure}
    \centering
    \includegraphics[width=0.4\textwidth]{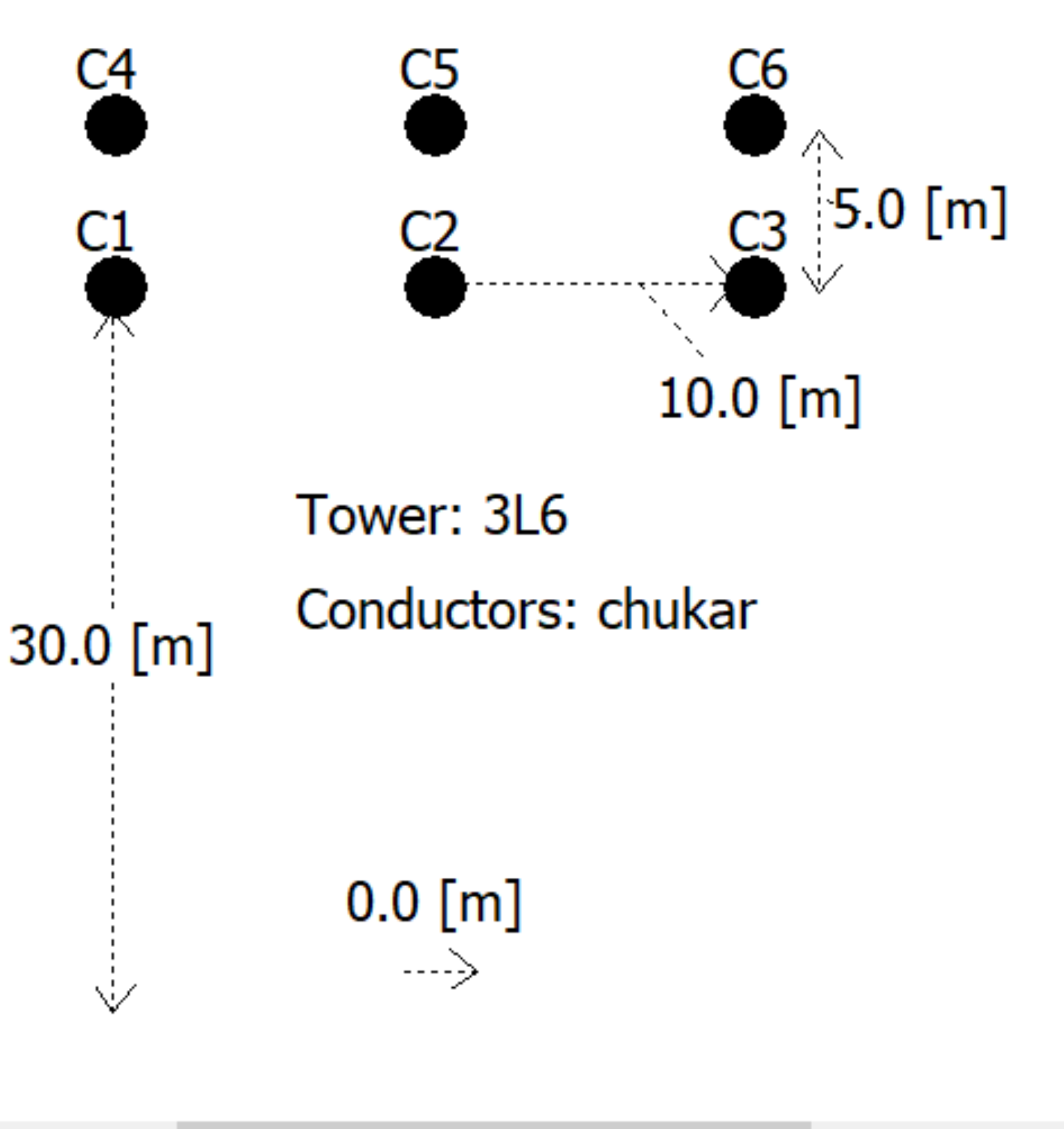}
    \caption[Transmission line model]{Transmission line model}
    \label{fig:tower}
\end{figure}

The model of a radial three-terminal MTdc symmetric monopole is shown in Fig.~\ref{fig:zones}. The system is equipped with hybrid dcCB at the MMC terminals. The dc transmission lines are designed as frequency dependent models having $6$ conductors with a vertical spacing of $5m$ and horizontal spacing of $10m$ between the conductors. The transmission line model has been shown in Fig.~\ref{fig:tower}.

\section{Methodology} \label{sec:Method}
In this section we briefly introduce and describe the single ended impedance based approach for fault location, then we introduce and discuss the ensemble based XGB approach for fault location.

\subsection{Single-ended Impedance Based Approach}

\begin{figure}[!ht]
    \centering
    \includegraphics[width=0.5\textwidth]{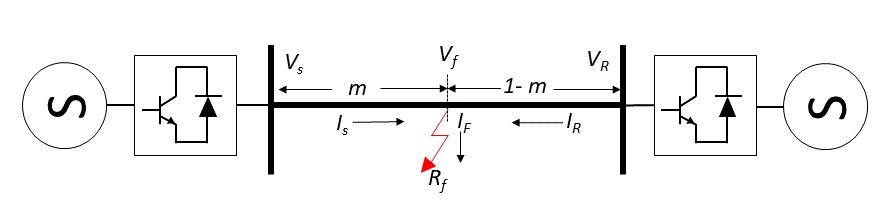}
    \caption{Single-ended impedance based fault location}
    \label{fig:single-endedfaultlocation}
\end{figure}

A single ended fault location involves the measurement of current and voltage at one end of the terminal as shown in Fig. \ref{fig:single-endedfaultlocation}. The calculated impedance of the transmission model along with other parameters are utilized for calculation of the terminal end voltage $V_S$ as Equation (\ref{Eqn:1}) \cite{Das2014}:

\begin{equation}
    V_S = mZ_lI_S + R_FI_F \label{Eqn:1}
\end{equation}
where $Z_l$ is the transmission line impedance, $m$ is the per unit location of the fault, $I_S$ is the current measured at one terminal, $R_F$ and $I_F$ are the fault resistance and fault current respectively. The main goal is to accurately estimate $m$.
The implementation of the single-ended impedance based fault location is simple but the accuracy is dependant on $R_F$. In real conditions estimating the fault resistance is challenging. Inaccurate estimation of the fault resistance proportionally decreases the accuracy of the method. To overcome the challenges posed by single-ended measurement based approaches we propose the utilization of an ensemble learning based approach for fault location.

\subsection{eXtreme Gradient Boosting (XGB)}\label{subsec:XGBoost}
XGB~\cite{russell2002artificial} is a supervised ensemble learning algorithm is an implementation of gradient boosting with pruning and regularization. This approach prevents overfitting and avoids cache misses by carefully and efficiently organizing memory thus allowing parallel computation. 
The accuracy of model fit is measured using a loss function, $L(\theta)$, such as mean squared error. For linear regression problems, $L(\theta)$ is
\begin{equation}\label{XGBMeanSquaredErr}
    L(\theta) = \sum_{i}(y_i-\hat{y}_i)^2
\end{equation} 
For logistic regression problems, $L(\theta)$ is
\begin{equation}\label{XGBLogisticLoss}
L(\theta) = \sum_{i}[{y_i}\ln(1+e^{-\hat{y}_i})+(1-y_i)\ln(1+e^{\hat{y}_i})]
\end{equation} 
\noindent where $y_i$ is the target variable and $\hat{y}_i$ is model's prediction.

The loss function can then be minimized with each successive iteration of the model using the formula 
\begin{equation}\label{eq:XGBFinalFunction}
    \hat{y}_i^{(m+1)}=\hat{y}_i^{(m)} - \gamma \frac{\partial{L}}{\partial{{\hat{y}_i}^{(m)}}}
\end{equation}
\noindent where $\gamma$ is the learning rate.

\begin{figure}
    \centering
    \includegraphics[width=0.5\textwidth]{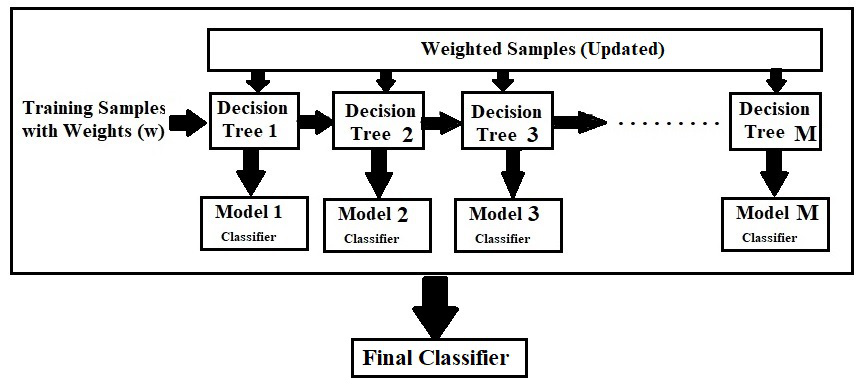}
    \caption{eXtreme Gradient Boosting Learning}
    \label{fig:xgboostModel}
\end{figure}

A conceptual representation model of ensemble learning is shown in Fig. \ref{fig:xgboostModel}.

\subsection{Data and Experimental Setup}\label{sec:Ex:ExperimentalSet}
As shown in Fig. \ref{fig:zones}, a 640 kV dc, radial MTdc transmission network was utilized for the single-ended fault location using XGB algorithm. The parameters for the test system are listed in Table \ref{Table 1}. Fault events were simulated on various sections of the dc transmission network using PSCAD/EMTdc. The fault locations were randomly chosen across the dc section of the network. Multiple scenarios were generated by varying the current limiting inductance (1 mH - 200 mH), fault resistance (0.01 $\Omega$ - 200 $\Omega$), change of loads in the transmission system (non-fault events). The current and voltage at the terminals were sampled at 1 kHz.

The generated data is standardized by removing the mean and scaling to the unit variance by computing the standard score (SS) using:

\begin{equation}\label{eq:stdscalar}
SS = \frac{x-u}{s}
\end{equation}
\noindent where $u$ and $s$ are mean and standard deviation of the training samples.

The processed data is then split into 7-folds and trained using XGB algorithm, where the objective is to minimize the loss function defined in Equation~(\ref{eq:XGBFinalFunction}). 

\section{Results and Discussion}\label{sec:Experiment}
This section describes application of machine learning algorithms to the data described in Section~\ref{sec:Ex:ExperimentalSet}. Results based on XGB algorithm is presented.

\subsection{Model Performance}\label{sec:Ex:Algo}
This section compares various machine learning algorithms including Ordinary Least Squares (OLS) regression~\cite{hutcheson2011ordinary}, Bayesian Regularization (BR)~\cite{burden2008bayesian}, Support Vector Regression (SVR)~\cite{drucker1997support}, k-Nearest Neighbor Regressor (KNNR)~\cite{maltamo1998methods}, Decision Tree (DTree)~\cite{russell2002artificial},XGB~\cite{russell2002artificial}, Gradient Boosting (GB)~\cite{russell2002artificial}, and Multi-layer Perceptron (MLP)~\cite{goodfellow2016deep} when trained on data presented in Section~\ref{sec:Ex:ExperimentalSet}.

\begin{figure}[!ht]
	\centerline{\includegraphics[width=0.5\textwidth]{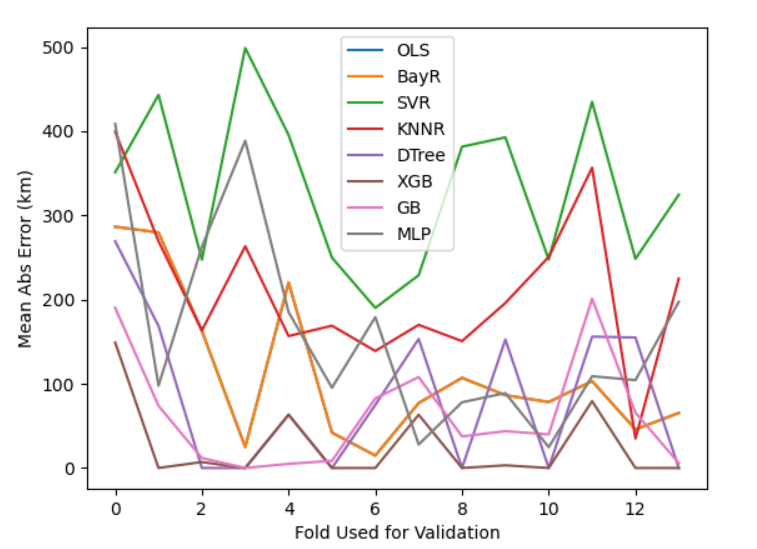}}
	\caption{Model performance }\label{fig:ModelFoldComparison}
\end{figure}

Fig.~\ref{fig:ModelFoldComparison} shows performance of models when trained on each fold of the data. The first half of Fig.~\ref{fig:ModelFoldComparison} between $0-6$ fold is measured on voltage data and the second half between $7-12$ fold is measured on current data.

\begin{figure}[!ht]
	\centerline{\includegraphics[width=0.5\textwidth]{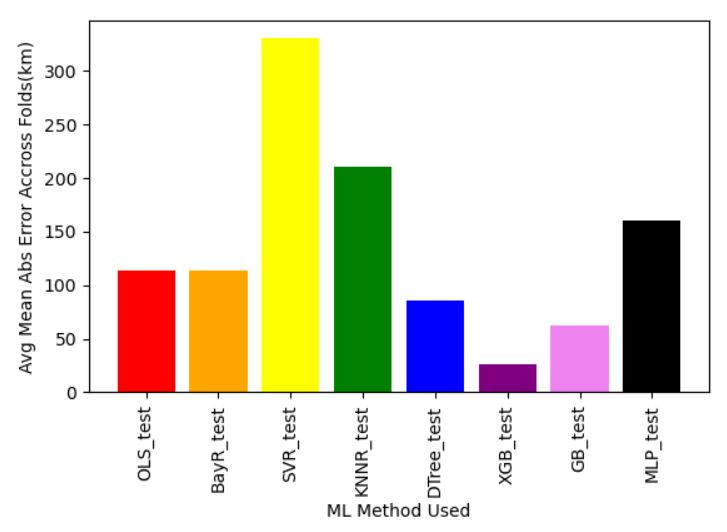}}
	\caption{Average of model's performance comparison}\label{fig:ModelComparison}
\end{figure}

\begin{figure*}[!ht]
	\centerline{\includegraphics[width=\textwidth]{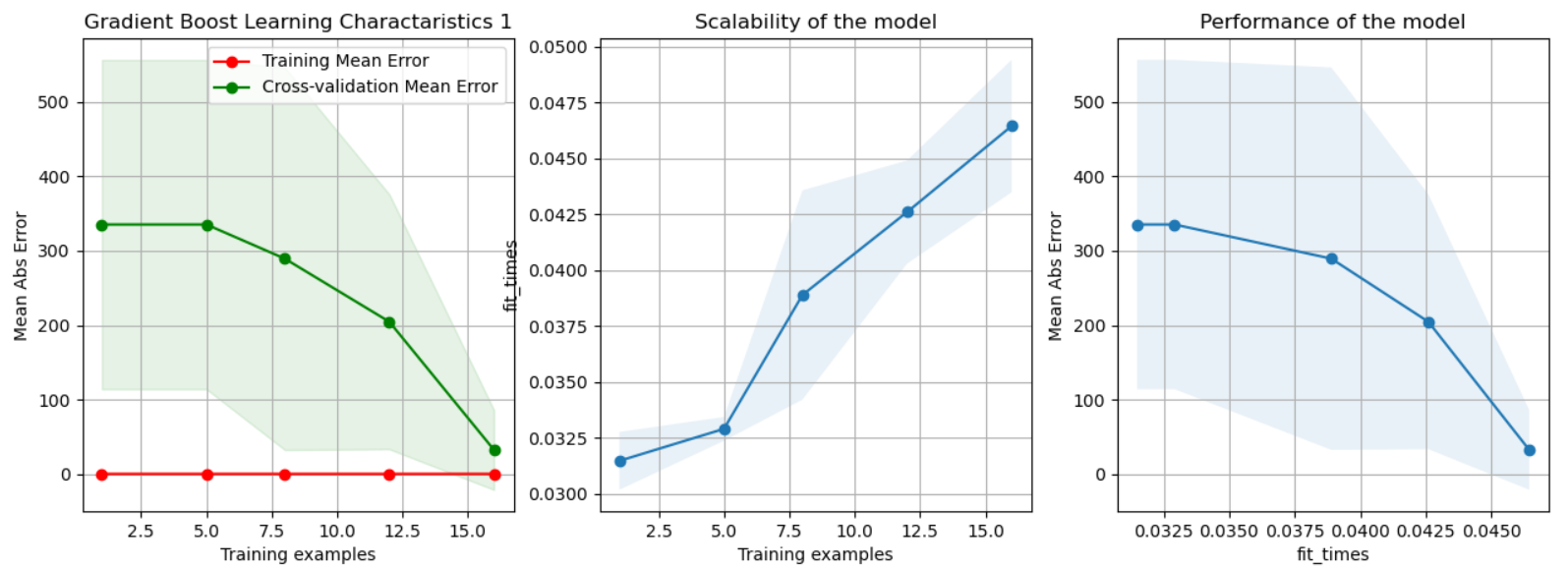}}
	\caption{Performance and scalability of XGB model on voltage data}\label{fig:XGBResults}
\end{figure*}

\begin{figure*}
	\centerline{\includegraphics[width=\textwidth]{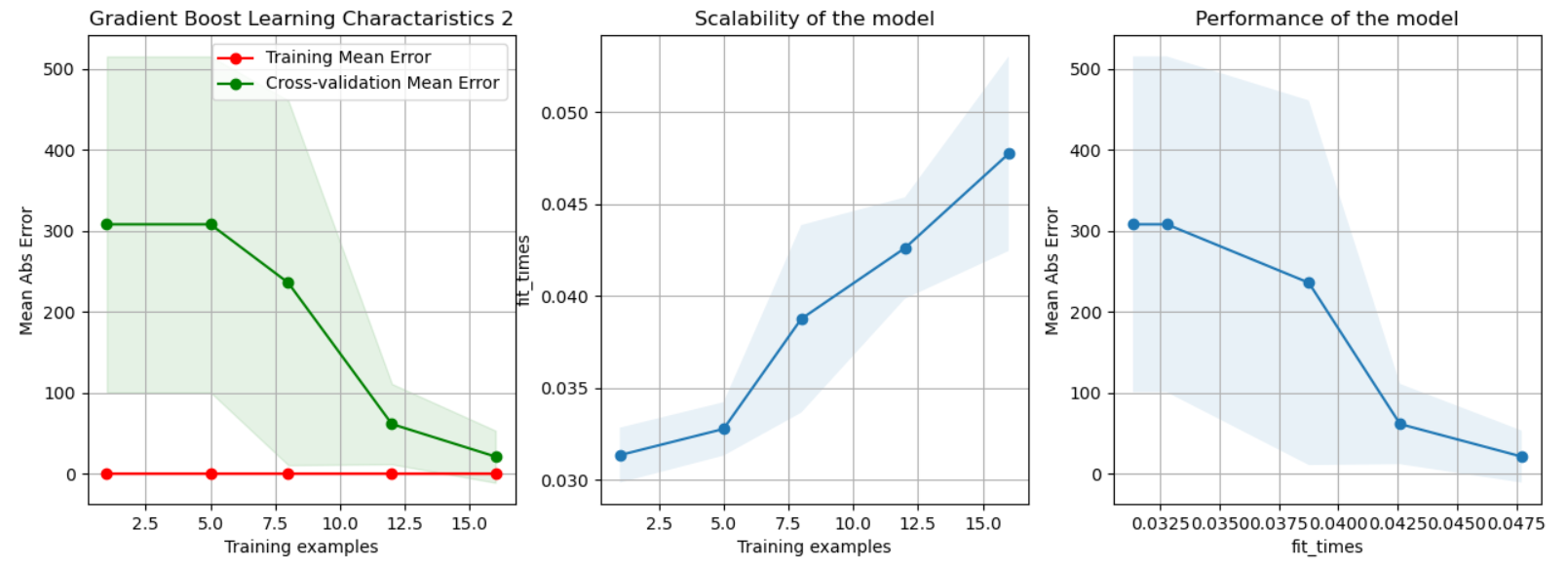}}
	\caption{Performance and scalability of XGB model on current data}\label{fig:XGBResults2}
\end{figure*}

\noindent In Fig.~\ref{fig:ModelFoldComparison}, the performance is measured as the mean absolute (MAE) value of the prediction errors in kilometers from the target location, defined as follows,
\begin{equation}
    MAE = \frac{\sum_{i=1}^{n}\left| \hat{x_i}-x_i\right|}{n}
\end{equation}
where, $\hat{x_i}$ is the predicted fault location, $x_i$ is the actual location and \textit{n} is the number of data points.

While all algorithms present a significantly high variation in the error, XGB and GB consistently maintain relatively low variation regardless of small individual fold sizes used in the study.

%High variation in the error can be seen due to small individual fold sizes of three data points each. 

Alternatively, Fig.~\ref{fig:ModelComparison} shows the model's average performance across all folds of both voltage and current samples by taking the MAE across the current and voltage folds.
The average error of XGB model is approximately $30$ compared to all other models, which are above $50$, thus achieving $66.67\%$ lesser error value.

\subsection{XGB Results}
Fig.~\ref{fig:XGBResults} shows assessment of XGB algorithm when applied to voltage data described in Section~\ref{sec:Ex:ExperimentalSet}. The first graph in Fig.~\ref{fig:XGBResults} shows mean absolute error during learning/training and validation sets across folds when compared to the number of training samples/examples given to the model. The second graph shows the time taken for each amount of examples to train, and third graph shows performance over time rather than the performance per example.

\noindent In Fig.~\ref{fig:XGBResults}, the first subplot shows a decreasing mean error as more number of samples indicating better learning of data by XGB. The second subplot shows linear increase in training time, which demonstrates the computing efficiency. The third subplot shows the average error over time, which supports results shown in the first subplot.

Similarly, Fig.~\ref{fig:XGBResults2} assessment of XGB algorithm when applied to current data described in Section~\ref{sec:Ex:ExperimentalSet}.

\section{Conclusion} \label{sec:Conclusion}
In this paper, we proposed a single ended fault location technique using XGB, ensemble learning method. The single-ended post-fault voltage and current measurements are used as input to the XGB model. Several machine learning algorithms were utilized to support the selection of XGB algorithm. Results suggested that XGB achieved a consistently better accuracy compared to other algorithms despite the small folds and small sample size of the training data provided. Mean average error between estimating the fault location decreased from $300$ to a value closer to zero. XGB also demonstrated linear computing time with the increase in training samples.

% \section*{Acknowledgment}

\bibliographystyle{IEEEtran}
\bibliography{APAP2021.bib}

\end{document}